\documentclass[]{aipproc}
\layoutstyle{6x9}

\usepackage{amsmath}
\usepackage{units}
\usepackage{overpic}

\newcommand{\dd}{\ensuremath{\mathrm{d}}}

\newcommand{\Pgg}{\ensuremath{\mathrm{\gamma}}}
\newcommand{\Pp}{\ensuremath{\mathrm{p}}}
\newcommand{\Pgpz}{\ensuremath{\mathrm{\pi^0}}}
\newcommand{\Pgh}{\ensuremath{\mathrm{\eta}}}

\begin{document}

\title{Measurement of the polarization observables\linebreak[1] T, P, H in pseudoscalar meson photoproduction with the CBELSA/TAPS experiment}


\author{J. Hartmann \\for the CBELSA/TAPS collaboration}{
  address={Helmholtz-Institut f\"ur Strahlen- u. Kernphysik, Universit\"at Bonn, Germany}
}

\begin{abstract}
The CBELSA/TAPS experiment with a longitudinally or transversely polarized target and an energy tagged, linearly or circularly polarized photon beam allows the measurement of a large set of polarization observables. Due to its good energy resolution and high detection efficiency for photons, as well as the nearly complete solid angle coverage, it is ideally suited for the measurement of neutral mesons decaying into photons.

Preliminary results for the target asymmetry $T$, the recoil polarization $P$, and the double polarization observable $H$ are shown for $\pi^{0}$ and $\eta$ photoproduction off the proton.
\end{abstract}

\maketitle

\section{Introduction}
In order to extract the contributing resonances in photoproduction experiments, partial wave analyses need to be performed. A complete experiment is required to determine the contributing amplitudes. This involves the measurement of single and double polarization observables.

For single pseudoscalar meson photoproduction using a linearly polarized photon beam and a transversely polarized target, the cross section can be written in the form
\begin{equation}
\frac{\dd\sigma}{\dd\Omega} = \left(\frac{\dd\sigma}{\dd\Omega}\right)_0 \left[1-P_{\Pgg}\Sigma\cos(2\phi) - P_xP_{\Pgg}H\sin(2\phi) - P_y\left(P_{\Pgg}P\cos(2\phi)-T\right)\right]
\end{equation}
where $\left(\frac{\dd\sigma}{\dd\Omega}\right)_0$ is the unpolarized cross section, $\Sigma$, $T$, $P$, and $H$ are the occurring polarization observables \cite{barker:1975}, $P_{\Pgg}$ the degree of linear photon polarization, and $\phi$ the azimuthal angle of the photon polarization plane with respect to the reaction plane. $P_x$ and $P_y$ are the degrees of target polarization in the direction of the produced meson and perpendicular to that, respectively.

\section{Data analysis and preliminary results}
The data presented has been obtained with the CBELSA/TAPS experiment at ELSA \cite{hillert:2006}. The detector system consists of two electromagnetic calorimeters, the Crystal Barrel \cite{aker:1992} and the MiniTAPS detector \cite{novotny:1991}, together covering the polar angle range from $1^\circ$ to $156^\circ$ and the full azimuthal angle. For charged particle identification, a three-layer scintillating fiber detector \cite{suft:2005} surrounding the target, and plastic scintillators in forward direction were used. The frozen spin butanol target \cite{bradtke:1999} was operated with a superconducting saddle coil providing a homogeneous magnetic field perpendicular to the beam direction, reaching an average target polarization of $\unit[74]{\%}$. Data has been taken with two opposite settings of the target polarization direction (named $\uparrow$ and $\downarrow$).
%For both settings, part of the data was taken with the target polarization parallel and antiparallel to the transverse holding field, in order to minimize systematic effects of the magnetic field on the recoiling proton.

For this analysis, a data set obtained with a primary electron energy of $\unit[3.2]{GeV}$ was used. The energy tagged photon beam was linearly polarized by means of coherent bremsstrahlung \cite{elsner:2009} with a maximum polarization of $\unit[65]{\%}$ at $E_\gamma = \unit[830]{MeV}$. Two perpendicular settings of the polarization plane were used (named $\parallel$ and $\perp$).

The data sample was selected for events with three distinct calorimeter hits, with one of them charged and two uncharged. Further kinematic cuts to ensure momentum conservation were applied, and the $\vec{\Pgg}\,\vec{\Pp} \to \Pp \Pgpz$ and $\vec{\Pgg}\,\vec{\Pp} \to \Pp \Pgh$ events were selected by applying a cut on the $\Pgg\Pgg$ invariant mass. This results in a final event sample containing a total of $1.4$ million $\Pp\Pgpz$ and $140000$ $\Pp\Pgh$ events with a background contribution of $<\unit[1]{\%}$ and $<\unit[5]{\%}$, respectively. The selected events for each of the four combinations of beam and target polarization directions were normalized w.r.t. the number of events and average polarization.

\begin{figure}
\begin{overpic}[width=0.33\textwidth]{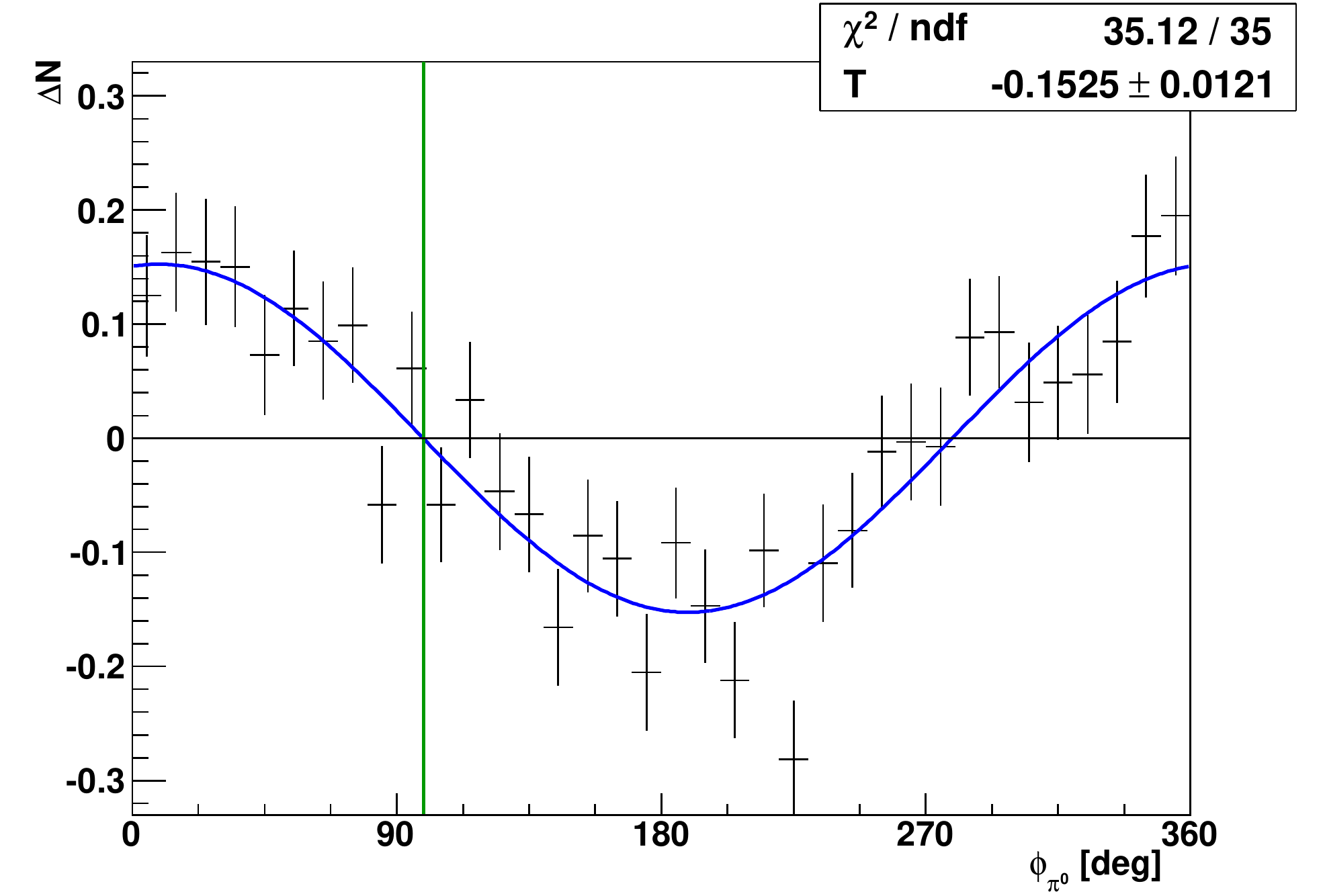}
\put(10,6.5){\includegraphics[width=0.27\textwidth]{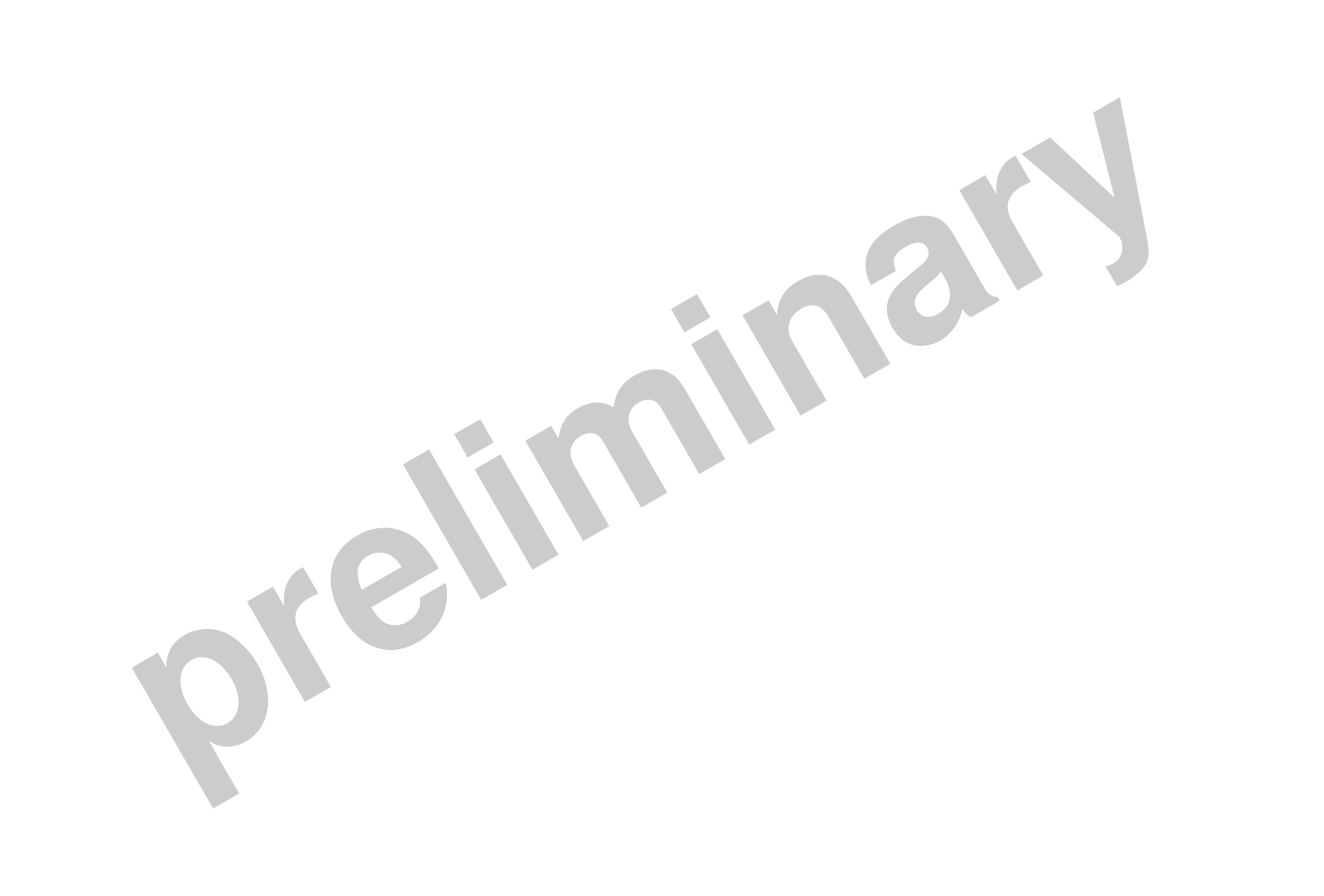}}
\put(31,2){\begin{tiny}$\beta$\end{tiny}}
% \put(75,1){\begin{tiny}$\phi_{\Pgpz} [^\circ]$\end{tiny}}
\put(10,65){\begin{tiny}$E_{\Pgg} = \unit[800]{MeV}$\end{tiny}}
\put(35,65){\begin{tiny}$\cos\theta = 0.25$\end{tiny}}
\put(0,65.5){\begin{footnotesize}(a)\end{footnotesize}}
\end{overpic}
\begin{overpic}[width=0.33\textwidth]{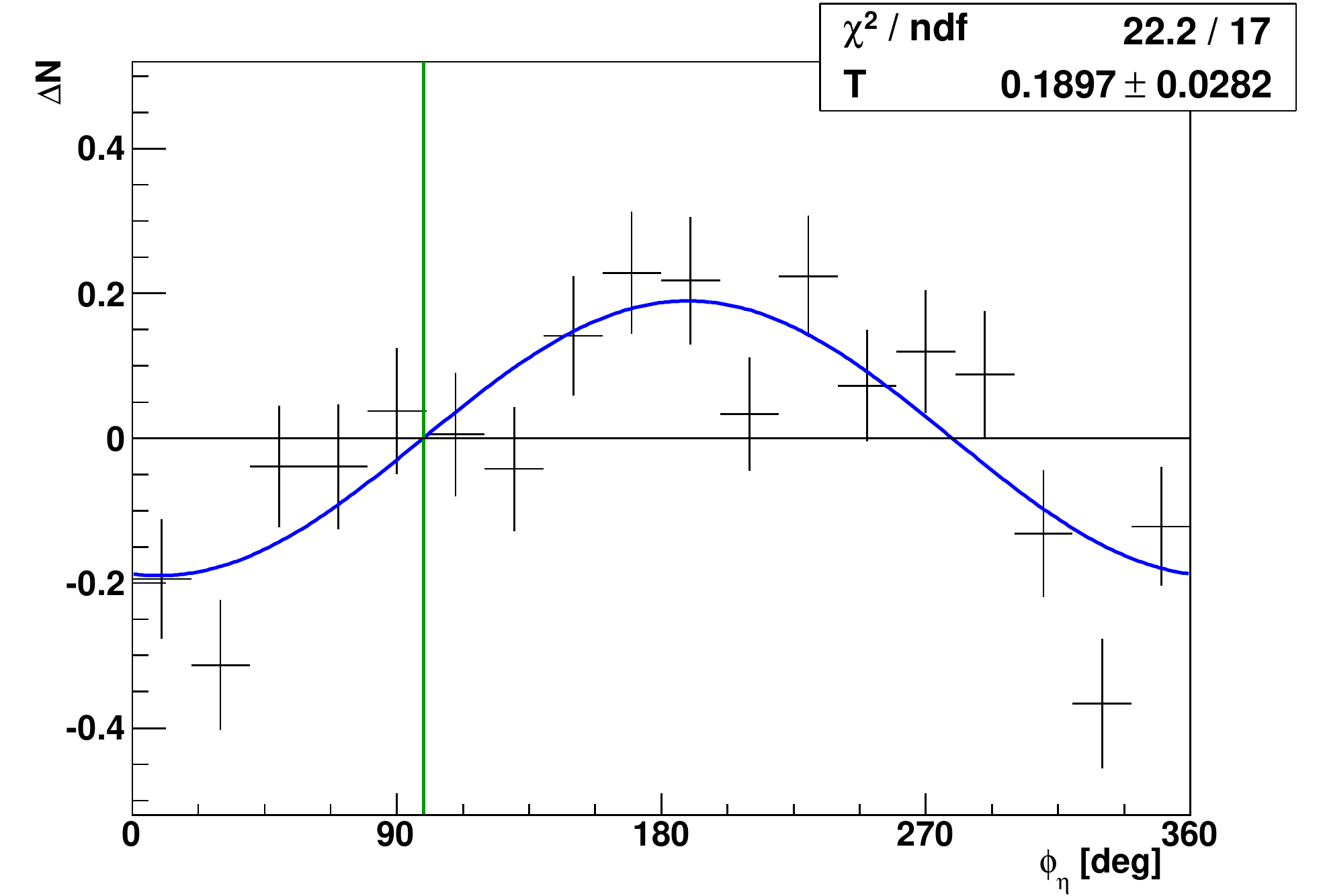}
\put(10,6.5){\includegraphics[width=0.27\textwidth]{preliminary}}
\put(31,2){\begin{tiny}$\beta$\end{tiny}}
% \put(75,1){\begin{tiny}$\phi_{\Pgh} [^\circ]$\end{tiny}}
\put(10,65){\begin{tiny}$E_{\Pgg} = \unit[825]{MeV}$\end{tiny}}
\put(35,65){\begin{tiny}$\cos\theta = 0.25$\end{tiny}}
\put(0,65.5){\begin{footnotesize}(b)\end{footnotesize}}
\end{overpic}
\begin{overpic}[width=0.33\textwidth]{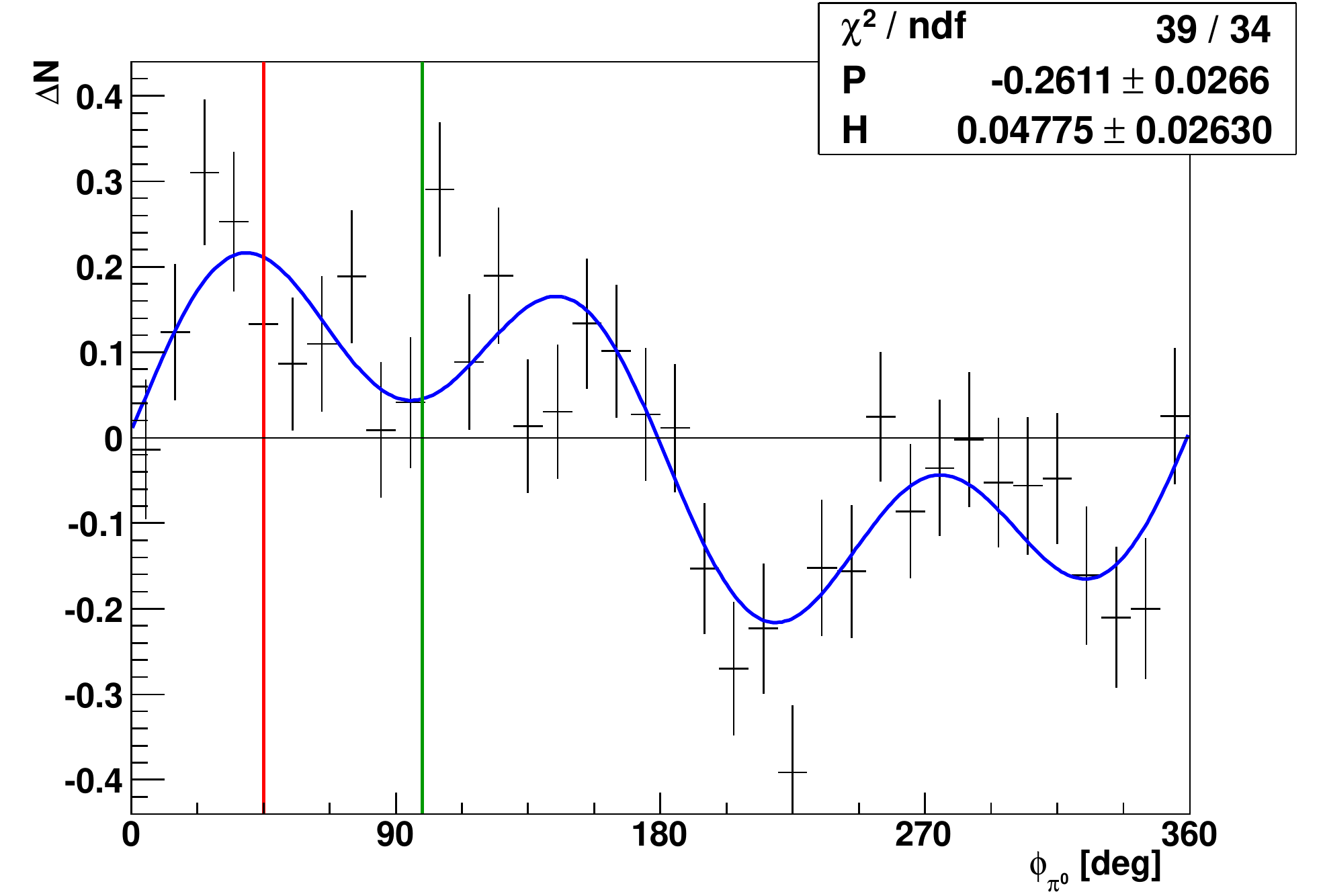}
\put(10,6.5){\includegraphics[width=0.27\textwidth]{preliminary}}
\put(18.5,2.2){\begin{tiny}$\alpha$\end{tiny}}
\put(31,2){\begin{tiny}$\beta$\end{tiny}}
% \put(75,1){\begin{tiny}$\phi_{\Pgpz} [^\circ]$\end{tiny}}
\put(10,65){\begin{tiny}$E_{\Pgg} = \unit[800]{MeV}$\end{tiny}}
\put(35,65){\begin{tiny}$\cos\theta = 0.25$\end{tiny}}
\put(0,65.5){\begin{footnotesize}(c)\end{footnotesize}}
\end{overpic}
\label{fig:phidistr}
\caption{Examples for measured $\phi$-distributions used to extract the target asymmetry $T$ in $\Pgpz$ (a) and $\Pgh$ photoproduction (b), and $P$ and $H$ in $\Pgpz$ photoproduction (c).}
\end{figure}

%\subsection{Target asymmetry $T$}
The target asymmetry $T$ can be determined using
\begin{equation}
\Delta N(\phi) = \frac{1}{f \cdot P_t} \cdot \frac{N_\uparrow-N_\downarrow}{N_\uparrow+N_\downarrow} = T \cdot \sin(\beta-\phi) ;\quad f(E_{\Pgg},\theta) = \frac{N_{butanol} - N_{carbon}}{N_{butanol}}
\end{equation}
with average target polarization $P_t$, $\beta=99^\circ$ being the direction of the target polarization in the $\uparrow$ setting, and the effective dilution factor $f$, which arises from the fact that not all protons in the butanol target are polarized. In order to determine $f$ for each $E_{\Pgg}$ and $\cos\theta$ bin, two additional data samples with unpolarized LH$_2$ and carbon targets were used. This data was normalized in such a way that the butanol data agrees with the sum of LH$_2$ and carbon data.
The target asymmetry is determined by a fit to the $\Delta N(\phi)$ distributions as shown by the examples in Fig.\ref{fig:phidistr}.

%\subsection{Recoil polarization $P$ and double polarization observable $H$}
Two additional observables are accessible by using linear beam polarization and a transversely polarized target. In addition to the observable $H$, one also has access to the recoil polarization $P$, without the difficulties associated with a direct measurement of the recoil proton polarization. In order to extract both observables from the data, all four combinations of beam and target polarization settings are used:
\begin{eqnarray}
\Delta N(\phi) &=&
\frac{1}{f \cdot P_{\Pgg} P_t} \cdot \frac{(N_{\perp\uparrow}-N_{\perp\downarrow})-(N_{\parallel\uparrow}-N_{\parallel\downarrow})}{(N_{\perp\uparrow}+N_{\perp\downarrow})+(N_{\parallel\uparrow}+N_{\parallel\downarrow})} \\
&=& P \sin(\beta-\phi)\cos(2(\alpha-\phi)) + H \cos(\beta-\phi)\sin(2(\alpha-\phi)) \notag
\end{eqnarray}
with average beam polarization $P_{\Pgg}$ and $\alpha=45^\circ$ being the direction of the polarization plane in the $\parallel$ setting. Again, the observables can easily be determined by a fit to the $\Delta N(\phi)$ distributions, as shown by the example in Fig.\ref{fig:phidistr}.

%\section{Preliminary results}
\begin{figure}
\includegraphics[width=\textwidth]{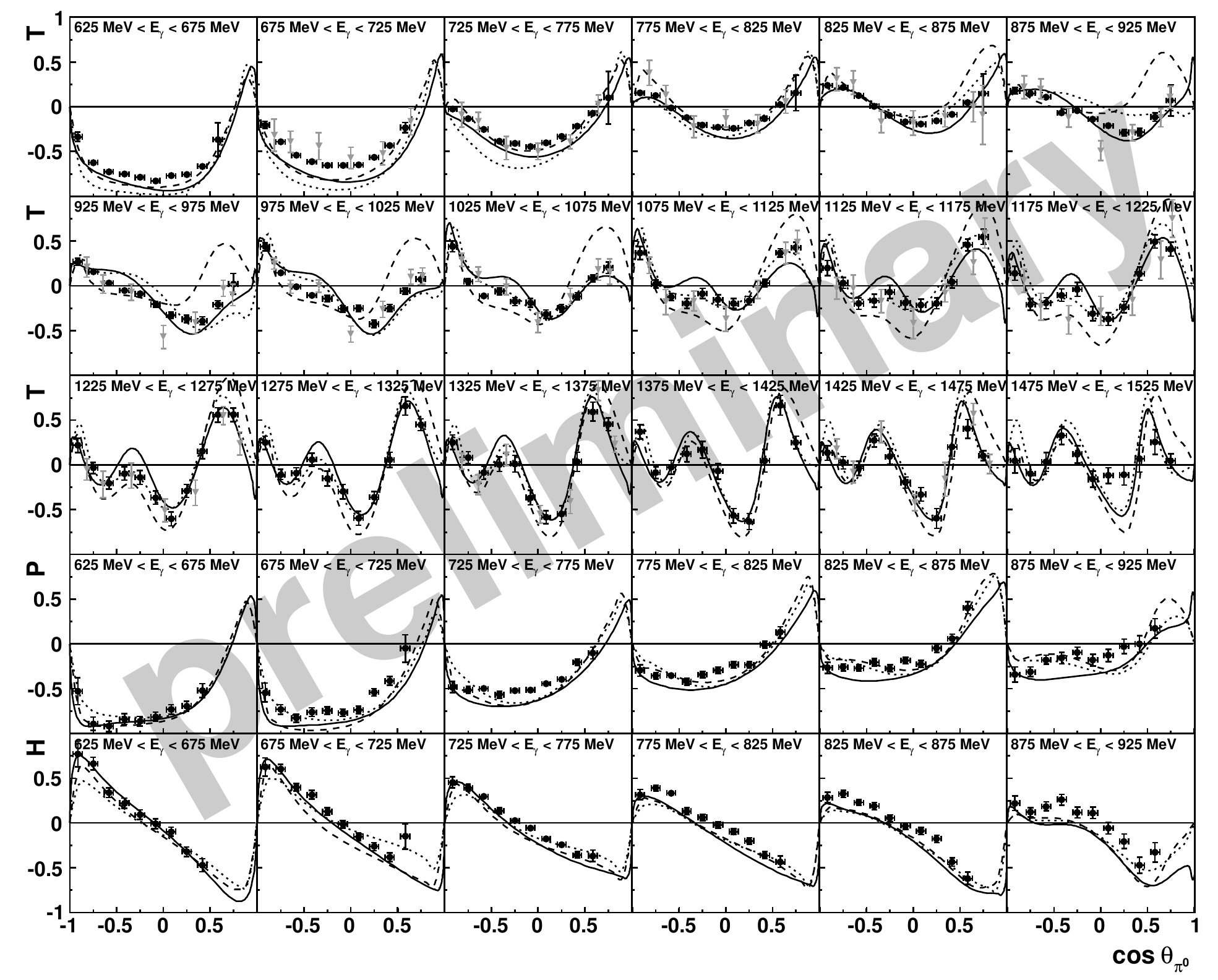}
\caption{Preliminary distributions of $T$ (top 3 rows), $P$ (4th row), and $H$ (bottom row) for the reaction $\vec{\Pgg}\,\vec{\Pp}\to\Pp\Pgpz$ (black) as a function of the $\Pgpz$ CMS angle, compared to previous measurements of $T$ \cite{booth:1977} (gray) and the predictions of the BnGa \cite{bnga} (solid), SAID \cite{said:2009} (dotted), and MAID \cite{maid:2007} (dashed) analyses.}
\label{fig:pi0}
\end{figure}
\begin{figure}
\includegraphics[width=\textwidth,trim=0cm 0cm 0cm 8.66cm,clip]{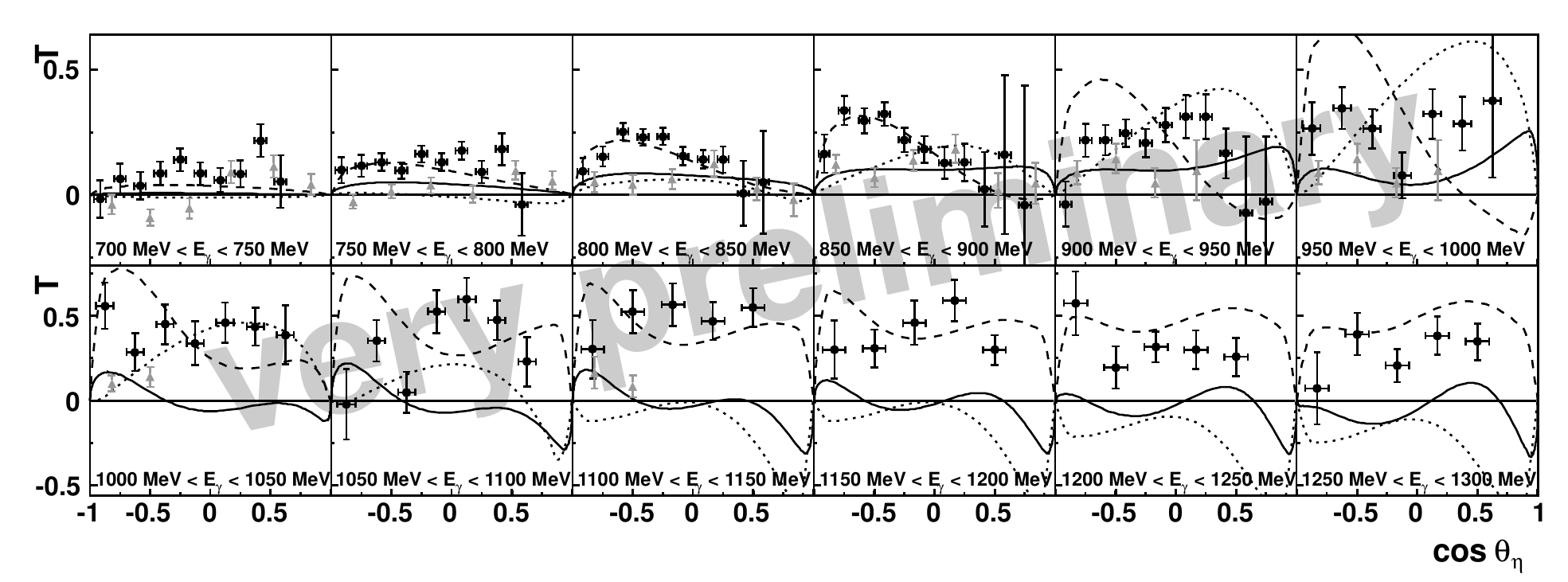}
\caption{Very preliminary distributions of $T$ for the reaction $\Pgg\,\vec{\Pp}\to\Pp\Pgh$ (black) as a function of the $\Pgh$ CMS angle, compared to previous measurements \cite{bock:1998} (gray) and the predictions of the BnGa \cite{bnga} (solid), SAID \cite{said:2009} (dotted), and MAID \cite{maid:2007} (dashed) analyses.}
\label{fig:eta}
\end{figure}

Preliminary results for the observables $T$, $P$, and $H$ are shown in Fig. \ref{fig:pi0} for $\Pgpz$ photoproduction. The shown error bars so far only include statistical uncertainties. The agreement with previous measurements of $T$ performed at the Daresbury $\unit[5]{GeV}$ electron synchrotron \cite{booth:1977} is quite good.

Very perliminary results for the target asymmetry $T$ in $\Pgh$ photoproduction are shown in Fig. \ref{fig:eta}. The results seem to be inconsistent with previous measurements done at PHOENICS \cite{bock:1998}, but a detailed study of the systematic uncertainties needs to be done before a final conclusion can be drawn.

\section{Summary}
Data has been taken with the CBELSA/TAPS experiment using the newly developed transversely polarized target and a linearly polarized photon beam. The preliminary results show the excellent quality of the data for the reaction $\vec{\Pgg}\,\vec{\Pp} \to \Pp\Pgpz$. Further measurements are planned to increase the statistics to investigate other reactions and higher energies.
Together with the measurements with a longitudinally polarized target and both linearly \cite{thiel:2010} or circularly polarized photon beams this is an important step towards a complete experiment and will provide further constraints for the partial wave analysis.

\begin{theacknowledgments}
%We thank the technical staff of ELSA and the participating institutions for their invaluable contributions to the experiment.
This work was supported by the \emph{Deutsche Forschungsgemeinschaft} within SFB/TR-16.
\end{theacknowledgments}

\bibliographystyle{aipproc}   % if natbib is available

\bibliography{cb_hartmann}

\end{document}